\documentclass[12pt]{article}
\usepackage{amssymb,amsmath,epsfig}
\allowdisplaybreaks
\begin{document}
\title{\bf Quasi-normal Modes and Thermal Fluctuations of Charged Black Hole with Weyl Corrections}

\author{M. Sharif \thanks{msharif.math@pu.edu.pk} and Zunaira Akhtar
\thanks{zunairaakhtar.pu@gmail.com}\\
Department of Mathematics, University of the Punjab,\\
Quaid-e-Azam Campus, Lahore-54590, Pakistan.}
\date{}
\maketitle
\begin{abstract}
In this paper, we study thermodynamics, quasi-normal modes and
thermal fluctuations of a charged black hole with Weyl corrections.
We first obtain thermodynamic quantities such as Hawking
temperature, entropy, and heat capacity for non-rotating as well as
rotating versions of this black hole. We also evaluate temperature
through quantum tunneling mechanism which is exactly the same as
found through surface gravity. We then discuss the relation between
Davies's point and quasi-normal modes. Finally, we investigate the
effects of thermal fluctuations on the uncorrected thermodynamic
quantities. It is concluded that the logarithmic corrections
originated from thermal fluctuations make the system more unstable
for small BHs.
\end{abstract}

\section{Introduction}

Black hole (BH) as a thermodynamical object is one of the most
fascinating objects in gravitational physics. The laws of BH
mechanics are analogous to the laws of thermodynamics relating area
of event horizon to entropy, surface gravity to temperature and mass
to energy.  These analogies compelled Bekenstein to determine a
quantitative relation between area of event horizon and entropy of
BH. However, the proposed relation seemed to violate the second law
of thermodynamics as anything that goes into BH can never be
retrieved back. Thus, it is impossible to obtain thermal equilibrium
between the BH and thermal radiation. The studies of BH at quantum
level reveal that BHs emit subatomic particles named as Hawking
radiation which provide information about the BH geometry.
Therefore, the Bekenstein entropy relation needs to be corrected
which leads to the concept of thermal fluctuations and paved the way
of holographic principle \cite{1}.

Hawking \cite{2a} examined the existence of BH radiations as a
tunneling spectrum of particles which are generated in the form of
pairs either inside or outside the event horizon. If these particles
generated inside the BH, the positive energy particles tunnel away
from the horizon whereas for the reverse scenario, the negative
energy particles tunnel inwards the horizon. Hence for both cases,
the positive energy particles leave to infinity and appear in the
form of Hawking radiation while the negative energy particles are
absorbed by the BH and decrease its mass. Classically, a BH is
considered to be a stable object and becomes unstable due to quantum
tunneling phenomena. In literature, several theoretical methods have
been proposed to study the spectrum of Hawking radiation but the
quantum tunneling approach is a proficient one as it visualizes the
radiation source effectively \cite{3}-\cite{2c}.

Quasi-normal modes (QNMs) of BHs are solutions of the perturbation
equations which allow to distinguish between BHs and other stellar
structures. Vishveshwara \cite{102b}, as a pioneer, calculated QNMs
from the scattering of gravitational waves for Schwarzschild BH.
Jing and Pan \cite{102''} studied the connection between QNMs and
phase transition of Reissner-Nordstrom (RN) BH and found that the
real as well as imaginary parts of QNMs behave as oscillatory
functions of charge. He et al. \cite{102DD} analyzed QNMs of scalar
perturbation for charged Kaluza-Klein BH and derived a relation
between the Davies point and QNMs.

Konoplya and Zhidenko \cite{102DD'} studied various aspects of BH
perturbations such as decoupling of variables in the perturbation
equations, QNMs, gravitational stability and holographic
superconductors. They analyzed the observational possibilities for
detecting QNMs of BHs and discussed the eikonal regime of QNMs
frequencies because of its correspondence with null geodesics
\cite{102Dd'}. It is shown that this correspondence is guaranteed
for any stationary spherically symmetric asymptotically flat BH.
Breton et al. \cite{102DD''} studied QNMs of Born-Infeld-de Sitter
BH and employed null geodesic to study the QNMs frequencies at the
eikonal limit. Churilova \cite{102DD'''} deduced a general formula
for eikonal QNMs for the class of asymptotically flat spacetimes and
extended theories of gravity in the form of Schwarzschild eikonal
QNMs. Moreover, some other researchers observed the asymptotic
behavior of QNMs (highly damped modes) for BH solutions in Weyl
gravity \cite{102DD''''}. They examined the validity of the relation
between calculated QNMs and unstable circular null geodesics.

It is believed that null geodesics and radius of photon sphere
describe important facts about the spacetime structure which have
great connection with QNMs of compact objects. Ghaderi and
Malakolkalami \cite{102dd} studied null geodesics as well as
thermodynamic properties of Bardeen BH in the presence of
quintessential field. Using null geodesics and photon sphere, Wei
and Liu \cite{102c} proposed a relation between QNMs and Davies
point for de Sitter RN BH and found that angular velocity as well as
Lyapunov exponent correspond to the real and imaginary parts of
QNMs, respectively. Wei et al. \cite{102cc} examined the null
geodesics of a test particle in the equatorial plane for rotating
Kerr-anti-de Sitter (AdS) BH and investigated the relationship
between thermodynamic phase transition and unstable circular photon
orbit.

One of the important issues in the BH thermodynamics is the
consideration of thermal fluctuations. These fluctuations are due to
statistical perturbations in compact objects and become effective
for small BHs \cite{103a, 103b}. It is believed that the emission of
Hawking radiation reduce the size of BH and ultimately increases its
temperature, hence the impact of thermal fluctuations on BH geometry
cannot be denied. Faizal and Khalil \cite{103c} studied the effects
of statistical fluctuations on the thermodynamics of RN, Kerr and
charged AdS BHs. Pourhassan et al. \cite{28,28'} computed the
leading-order correction terms for modified Hayward BH and analyzed
its stability against thermal fluctuations. Jawad and Shahzad
\cite{28c} evaluated the corrected entropy as well as specific heat
for regular BHs. When we consider the corrections to entropy, it is
physically more acceptable that entropy and temperature governed by
the first law of BH thermodynamics should be corrected at the same
time. Moreover, the first-order corrections have been applied to RN
AdS and Kerr-Newman AdS BHs to check the effects of corrected
entropy on thermodynamic potentials \cite{28cc}.

Sinha \cite{28a} examined the effect of thermal fluctuations on AdS
Kerr-Newman BH and observed that the logarithmic terms appear due to
entropy-area relation. Upadhyay et al. \cite{28aa} studied the
contribution of logarithmic corrections to the stability of charged
BH thermodynamics and observed that first-order corrections has no
effect on phase transition. Haldar and Biswas \cite{28c.} examined
first-order corrections in entropy for charged Gauss-Bonnet BH and
computed thermodynamic potentials such as Helmholtz free energy,
enthalpy and Gibbs free energy. Pourhassan and Upadhyay \cite{28c..}
analyzed BH stability as well as phase transition against critical
points. Recently, Pradhan \cite{28c...} studied the criteria of
second-order phase transition for charged accelerating BH and
computed thermodynamic quantities under logarithmic corrections.

This paper is devoted to studying thermodynamic quantities, QNMs and
leading-order correction terms for charged BH with Weyl corrections.
The paper is organized as follows. In section \textbf{2}, we
calculate thermodynamic quantities for non-rotating as well as
rotating charged BH. We also evaluate temperature through quantum
tunneling approach. Section \textbf{3} explores relationship between
QNMs and Davies point. In section \textbf{4}, we investigate the
effects of thermal fluctuations on thermodynamic quantities.
Finally, we summarize the results in the last section.

\section{Thermodynamics}

This section is devoted to deriving the thermodynamic quantities
such as Hawking temperature, entropy and heat capacity for
non-rotating/rotating versions of charged BH. We also analyze
thermodynamic stability of the system through heat capacity.

\subsection{Non-rotating Charged BH with Weyl Corrections}
\begin{figure}\center
\epsfig{file=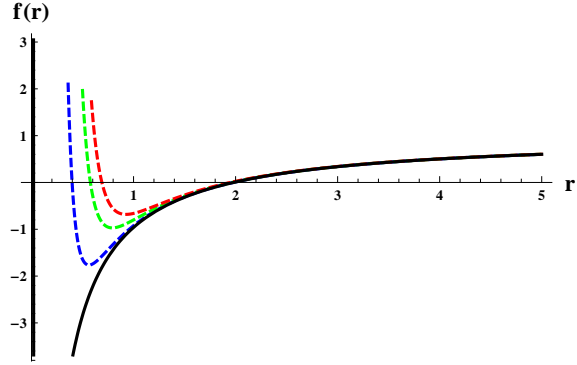,width=0.55\linewidth}\caption{Metric function
versus $r$ for $M=q=1$, and $\alpha=0$ (black), -0.2 (blue), -0.4
(green) and -0.6 (red).}
\end{figure}

The line element for a charged BH with Weyl corrections is given as
\cite{28c'}
\begin{equation}\label{1a}
ds^{2}=f(r)dt^{2}-\frac{1}{f(r)}dr^{2}-R(r)(d\theta^{2}+\sin^{2}\theta
d\phi^{2}),
\end{equation}
with
\begin{equation}\label{2}
f(r)=1-\frac{2 M}{r}+\frac{q^2}{r^2}-\frac{4 \alpha q^2}{3
r^4}+\frac{40\alpha M q^2}{9 r^5}-\frac{104 \alpha q^4}{45
r^6},\quad R(r)=r^{2}+\frac{4\alpha q^{2}}{9r^{2}},
\end{equation}
where $M$, $q$ and $\alpha$ represent the mass, charge and Weyl
coupling parameter, respectively. From the graphical analysis of
Figure \textbf{1}, we can see that negatively higher values of the
Weyl correction result in the appearance of naked singularities.
Moreover, $\alpha=0$ (no Weyl corrections) causes a negative
asymptote which shows that the solution has only one horizon. A
detailed analysis of the horizons and static limit surface has been
discussed in \cite{28c'}. It is noted that for $\alpha=0$, the above
metric reduces to the RN solution. The event horizon $(r_{+})$ of
the BH can be determined by setting $f(r_{+})=0$. For the considered
line-element, the explicit expression of BH horizon cannot be
obtained due to the presence of higher-order terms in the metric
potential. Setting $f(r_{+})=0$, the mass of the BH in terms of
$r_{+}$ is expressed as
\begin{equation}\label{3}
M=\frac{45 r_{+}^6+45 q^2 r_{+}^4-60 \alpha q^2 r_{+}^2-104 \alpha
q^4}{10 r_{+} \left(9 r_{+}^4-20 \alpha q^2\right)}.
\end{equation}
Hawking proposed that the radiation spectrum emitted from the BH
assists to determine its thermodynamic properties \cite{28c''}. In
BH physics, Hawking temperature has analogy with its surface gravity
$(\kappa=\frac{f^{'}(r_{+})}{2})$ as
\begin{equation*}
\kappa=\frac{81r_{+}^8 (q^{2}-r_{+}^{2})+16 \alpha^2 \left(26 q^6-15
q^4 r_{+}^2\right)+\alpha \left(576 q^2 r_{+}^6-396 q^4
r_{+}^4\right)}{18r_{+}^7 \left(-9 r_{+}^4+20 \alpha q^2\right)}.
\end{equation*}
The corresponding Hawking temperature $(T=\frac{\kappa}{2\pi})$ is
calculated as
\begin{equation}\label{5}
T=\frac{81r_{+}^8 (q^{2}-r_{+}^{2})+16 \alpha^2 \left(26 q^6-15 q^4
r_{+}^2\right)+\alpha \left(576 q^2 r_{+}^6-396 q^4
r_{+}^4\right)}{36 \pi r_{+}^7 \left(-9 r_{+}^4+20 \alpha
q^2\right)},
\end{equation}
whose graphical representation with respect to $\alpha$ is displayed
in Figure \textbf{2}. It is observed that the temperature decreases
for the larger values of $\alpha$ and $q$. The entropy in terms of
area law \cite{103a} can be defined as
\begin{equation}\label{6}
S=\frac{A}{4}=\pi r_{+}^{2}.
\end{equation}
\begin{figure}\center
\epsfig{file=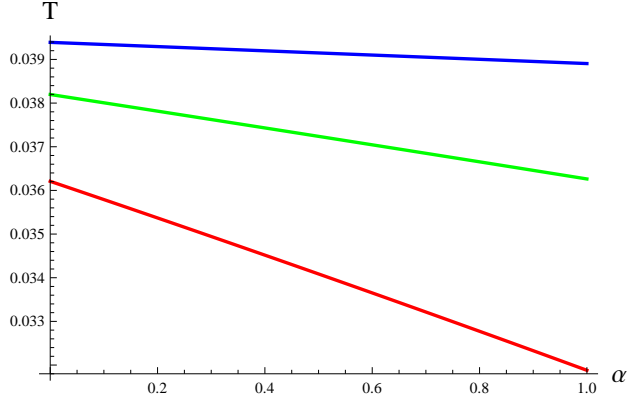,width=0.6\linewidth}\caption{Hawking
temperature versus $\alpha$ for $r_{+}=2$ and $q=0.2$ (blue), 0.4
(green) and 0.6 (red).}
\end{figure}
\begin{figure}\center
\epsfig{file=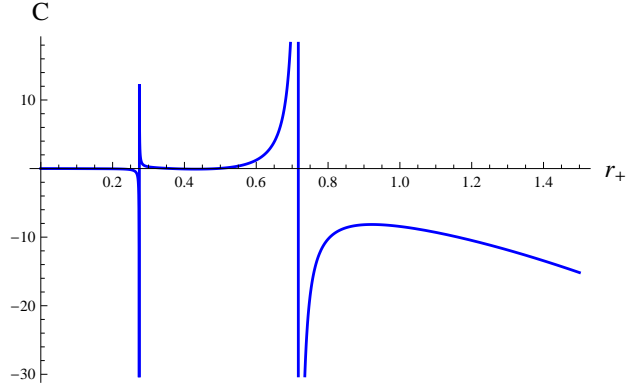,width=0.6\linewidth}\caption{Heat capacity versus
$r_{+}$ for $q=0.2$.}
\end{figure}

In order to investigate thermodynamic stability of BH, we evaluate
heat capacity $(C=T\frac{\partial S}{\partial T})$ as follows
\begin{eqnarray}\nonumber
C&=&-2 \pi r_{+}^2 \left(-9 r_{+}^4+20 \alpha q^2\right) \Big[81
r_{+}^8 (q^{2}-r_{+}^{2})+16 \alpha^2 \left(26 q^6-15 q^4
r_{+}^2\right)\\\nonumber&+&\alpha \left(576 q^2 r_{+}^6-396 q^4
r_{+}^4\right) \Big]\Big[320 \alpha^3 \left(182 q^8-75 q^6
r_{+}^2\right)\\\nonumber&+&144 \alpha^2 q^4 r_{+}^4\left(215
r_{+}^2-451 q^2\right)+324 \alpha q^2 r_{+}^8 \left(72 q^2-65
r_{+}^2\right)\\\label{7}&+&729 r_{+}^{12} \left(r_{+}^2-3
q^2\right)\Big]^{-1}.
\end{eqnarray}
It is known that BH is thermodynamical stable for positive values of
heat capacity whereas its negative values lead the system towards
instability \cite{28ci}.  Figure \textbf{3} shows that the heat
capacity diverges at two points $(r_{+}=0.27, 0.72)$ which leads to
the second-order phase transition. The heat capacity remains
positive around the first divergence point while it changes from
negative to positive at the second point. Thus, small BHs are
thermodynamically more stable as compared to the large ones.

Now we investigate the relation between Davies point and QNMs. Since
the explicit expression of BH horizon cannot be obtained due to the
presence of higher-order terms in the metric potential, we neglect
the higher-order terms and take the radial potential as follows
\begin{equation}\label{37}
f(r)=1-\frac{2 M}{r}+\frac{q^2}{r^2}-\frac{4 \alpha  q^2}{3 r^4}.
\end{equation}
Consequently, the temperature is
\begin{equation}\label{39}
T=\frac{8 \alpha q^2+3 M r_{+}^3-3 q^2 r_{+}^2}{6 \pi  r_{+}^5}.
\end{equation}
The corresponding heat capacity can be expressed as
\begin{equation}\label{40}
C=-\frac{2\pi r_{+}^2\left(8\alpha q^2+3r_{+}^2\left(M
r_{+}-q^2\right)\right)}{40\alpha q^2+6 M r_{+}^3-9 q^2 r_{+}^2}.
\end{equation}
The divergence point of heat capacity is known as Davies point which
measures a phase transition of the BH between thermodynamic stable
and unstable phases. In order to determine the divergence point of
heat capacity, the following identity must hold
\begin{equation}\label{41}
40\alpha q^2+6 M r_{+}^3-9 q^2 r_{+}^2=0.
\end{equation}
Here, it is difficult to compute the divergence point as $q$ cannot
be evaluated in terms of $M$ explicitly. Therefore, we plot the heat
capacity to determine its physical behavior as well as divergence
point. Figure \textbf{4} shows that BH is thermodynamically stable
and becomes unstable for $q>0.507$. Also, heat capacity diverges at
$q=0.55$ which indicates the first-order phase transition.
\begin{figure}\center
\epsfig{file=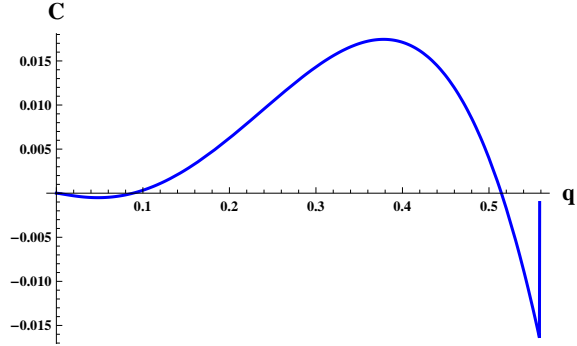,width=0.55\linewidth}\caption{Heat capacity
versus $q$ for $\alpha=0.0195$.}
\end{figure}

\subsection{Temperature through Quantum Tunneling}

Here we use quantum tunneling approach to obtain temperature for the
above mentioned BH. We only consider radial trajectories of
particles for which Eq.(\ref{1a}) reduces to
\begin{equation}\label{17}
ds^{2}=f(r)dt^{2}-\frac{1}{f(r)}dr^{2}.
\end{equation}
The Klein-Gordon equation with scalar field having mass $m_{\Phi}$
is
\begin{equation}\label{18}
\hbar^{2} \nabla\Phi-m^{2}_{\Phi}\Phi=0,
\end{equation}
where $\hbar$ is the Dirac constant. The corresponding D'Alembertian
operator turns out to be
\begin{equation}
\Box\Phi=-\frac{1}{f(r)}\ddot{\Phi}+f(r)\Phi^{''}+f^{'}(r)\Phi^{'}.
\end{equation}
This equation can be solved by applying Wentzel-Kramers-Brillouin
approximation which relates $\Phi$ with the action $I$ \cite{28c1}
\begin{equation}\label{19}
\Phi(t,r)=\exp\left[\frac{-i}{\hbar}I(t,r)\right].
\end{equation}
The corresponding Hamilton-Jacobi equation is
\begin{equation}\label{20}
\dot{I}^{2}-f^{2}(r)I^{'2}-m^{2}_{\Phi}f(r)=0,
\end{equation}
whose solution in terms of radiation energy $E$ and Hamilton
characteristic function $W(r)$ can be expressed as $I=-Et+W(r)$.
Here
\begin{equation}\label{21}
W_{\pm}(r)=\pm\int\frac{dr}{f(r)}\sqrt{E^{2}-f(r)m^{2}_{\Phi}},
\end{equation}
which represents spatial part of the action for particles going
inside $(W_{-})$ and outside $(W_{+})$ the BH.

We consider only the outward moving particles $W_{+}$ and use the
spatial metric as $d\sigma^{2}=\frac{dr^{2}}{f(r)}$. Applying the
near horizon approximation
$f(r)=f(r_{+})+\acute{f}(r_{+})(r-r_{+})+...$, we have
$\sigma=\frac{2\sqrt{r-r_{+}}}{\sqrt{\acute{f}(r_{+})}},$ where $0<
\sigma<\infty$. In this scenario, Eq.(\ref{21}) takes the form
\begin{equation}\label{23}
W(\sigma)=im_{\Phi}\int\sqrt{1-\frac{4E^{2}}{\sigma^{2}
m^{2}_{\Phi}\acute{f}^{2}(r_{+})}}d\sigma,
\end{equation}
where $W_{+}=W$. Integration of the above expression yields
$W(r)=\frac{2\pi i E}{\acute{f}(r_{+})},$ and hence
\begin{equation}\label{24a}
I=\frac{2\pi i }{\acute{f}(r_{+})}E+\text{(real contribution)}.
\end{equation}
The tunneling probability of outgoing and incoming particles across
the horizons are defined as \cite{28c2}
\begin{equation}\label{25}
\text{Prob [out]}=\exp[-2\text{Im}W_{+}],\quad \text{Prob
[in]}=\exp[-2\text{Im}W_{-}].
\end{equation}
Here, the imaginary part of the action is same for both the incoming
and outgoing solutions so they will cancel out the effect of each
other. Thus, the tunneling probability for particles can be
expressed as
\begin{equation}\label{26}
\Gamma\propto\frac{\text{Prob[out]}}{\text{Prob[in]}}=\frac{\exp[-2\text{Im}W_{+}]}
{\exp[-2\text{Im}W_{-}]}= \exp(-4\text{Im}W_{+}),
\end{equation}
through value of $W(r)$, gives rise to
\begin{equation}\label{27}
\Gamma=\exp\Big[-\frac{4\pi
E}{\acute{f}(r_{+})}\Big]=\exp[-\frac{E}{T}].
\end{equation}
Comparing $\Gamma=\exp(-2\text{Im}I)$ with Boltzmann factor
(\ref{27}), we have
\begin{equation}\nonumber
T_{t}=\frac{E}{2\text{Im}I}=\frac{\acute{f}(r_{+})}{4\pi}.
\end{equation}
This is exactly the same as that found through surface gravity. It
is pointed out that there exists a correlation between emitted
particles and leaked information which resolves the information loss
paradox, i.e., $\Gamma=\exp(-2\text{Im}I)=e^{\Delta S}$, where
$\Delta S$ is the difference between final and initial values of the
BH entropy. We plot the solution (14) with respect to horizon radius
to understand the behavior of scalar field. Figure \textbf{5}
represents that the scalar field behaves as damped periodic
oscillator which vanishes at horizon. It is indeed a bound state
which vanishes at infinity \cite{4a}.
\begin{figure}\center
\epsfig{file=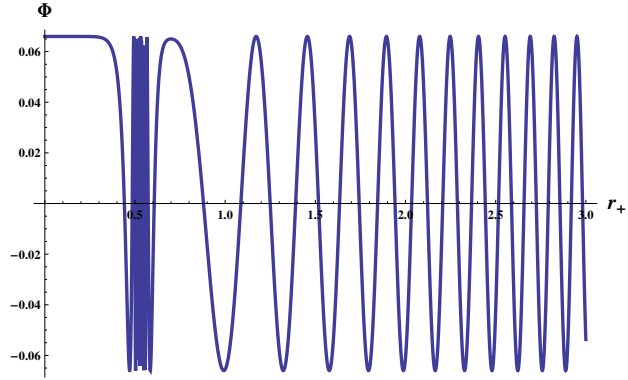,width=0.6\linewidth}\caption{Scalar field
versus $r_{+}$ for $m=t=1$ and $E=0.4$, $q$=0.2.}
\end{figure}

A tachyonic particle is a hypothetical massive particle that always
moves faster than the speed of light. It is observed that scalar
field is treated as a quantum field and an elementary particle is
described as an excitation near the minimum of the scalar potential.
The Taylor expansion of the scalar field near the minimum implies
that the coefficient of the quadratic term is always positive which
means that such a field is not tachyonic. However, if the Taylor
expansion near a maximum is observed, then the coefficient of the
quadratic term will be negative which leads the system towards
tachyonic field \cite{3a}. Comparing the derived results with
uncharged AdS BH \cite{4}, it is found that for $q=0$, all the
results reduce to uncharged scenario. The potential behaves as the
function of $ae^{-br^{2}}$ which corresponds to the effective
potential of tachyon field. It is found that massless scalar field
behaves as tachyon field in the absence of dilaton field background.
Hence, it depicts the same behavior as found in uncharged AdS BHs.

\subsection{Rotating Charged BH with Weyl Corrections}

The metric for rotating charged BH with Weyl corrections is defined
as \cite{28c'}
\begin{eqnarray}\nonumber
ds^{2}&=&\frac{F(r,\theta)}{\Sigma(r,\theta)}dt^{2}+2\Big[1-\frac{F(r,\theta)}
{\Sigma(r,\theta)}\Big]a\sin^{2}\theta dtd\phi\\\nonumber
&-&\Big[\frac{\Sigma(r,\theta)\Sigma_{1}(r,\theta)}{F(r,\theta)
\Sigma_{1}(r,\theta)+a^{2} \sin^{2}\theta
\Sigma(r,\theta)}\Big]dr^{2}-\Sigma_{1}(r,\theta)d\theta^{2}\\\label{8}
&-&\frac{\sin^{2}\theta}{\Sigma(r,\theta)}\Big[\Sigma(r,\theta)
\Sigma_{1}(r,\theta)+a^{2}\sin^{2}\theta(2\Sigma(r,\theta)-F(r,\theta))\Big]d\phi^{2},
\end{eqnarray}
such that
\begin{eqnarray}\nonumber
F(r,\theta)&=&r^2+a^2 \cos^{2}\theta -2 M r+q^2\\\nonumber
&-&\frac{4 \alpha q^2 }{3 \left(r^2+a^2
\cos^{2}\theta\right)}\left(1-\frac{50 M r-26 q^2}{15 \left(r^2+a^2
\cos^{2}\theta\right)}\right),\\\nonumber
\Sigma_{1}(r,\theta)&=&r^{2}+a^{2}\cos^{2}\theta+\frac{4\alpha
q^{2}}{9(r^{2}+a^{2}\cos^{2}\theta)},\\\label{9}
\Sigma(r,\theta)&=&r^{2}+a^{2}\cos^{2}\theta,
\end{eqnarray}
where $a$ is the rotation parameter. It is noted that the
line-element (\ref{1}) can be retrieved by setting $a=0$. For
simplification, we assume
\begin{equation}\label{10}
f(r,\theta)=\frac{F(r,\theta)}{\Sigma(r,\theta)}.
\end{equation}
Inserting the values of $F(r,\theta)$ and $\Sigma(r,\theta)$ in the
above expression leads to
\begin{equation}\label{11}
f(r,\theta)=1+\frac{q^2-2 M r}{a^2+r^2}-\frac{4 \alpha q^2 \left(15
a^2+5 r (3 r-10 M)+26 q^2\right)}{45 \left(a^2+r^2\right)^3}.
\end{equation}
The corresponding mass can be expressed as
\begin{equation}\label{12}
M=\frac{45 \left(a^2+r_{+}^2\right)^2 \left(a^2+q^2+r_{+}^2\right)-4
\alpha q^2 \left(15 a^2+26 q^2+15 r_{+}^2\right)}{90 r_{+}
\left(a^2+r_{+}^2\right)^2-200 \alpha  q^2 r_{+}}.
\end{equation}
The Hawking temperature for the rotating charged BH is given by
\begin{eqnarray}\nonumber
T&=&\Big[-405 \left(a^2+r_{+}^2\right)^4
\left(a^2+q^2-r_{+}^2\right) -80 \alpha ^2 q^4 \left(15 a^2+26
q^2-15 r_{+}^2\right)\\\nonumber&+&36 \alpha q^2
\left(a^2+r_{+}^2\right) \left(40 a^4+a^2 \left(51 q^2-40
r_{+}^2\right)+55 q^2 r_{+}^2-80
r_{+}^4\right)\Big]\\\label{14}&\times& \Big[180 \pi r_{+}
\left(a^2+r_{+}^2\right)^3 \left(9 \left(a^2+r_{+}^2\right)^2-20
\alpha q^2\right)\Big]^{-1}.
\end{eqnarray}
\begin{figure}\center
\epsfig{file=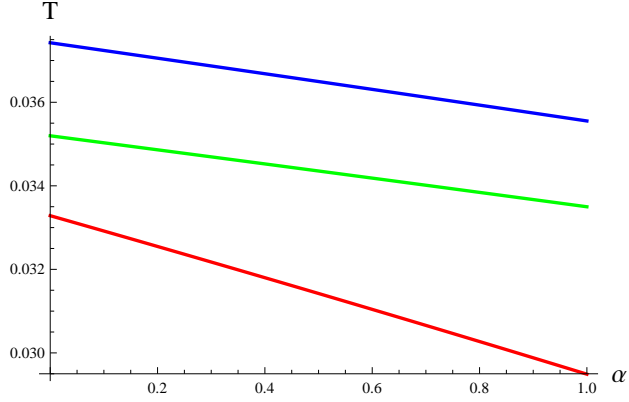,width=0.6\linewidth}\caption{Hawking
temperature versus $\alpha$ for $r_{+}=2$, $a=0.2$ and $q$ = 0.2
(blue), 0.4 (green), 0.6 (red).}
\end{figure}

Figure \textbf{6} indicates the decreasing behavior of Hawking
temperature with respect to charge and Weyl coupling parameter. In
this scenario, the entropy is modified as follows \cite{28c3}
\begin{equation*}\label{15}
S=\pi( r_{+}^{2}+a^{2}).
\end{equation*}
The corresponding heat capacity takes the form
\begin{eqnarray}\nonumber
C&=&\Big[2 \pi  \left(a^2+r_{+}^2\right) \left(9 r_{+}
\left(a^2+r_{+}^2\right)^2-20 \alpha  q^2 r_{+}\right)^2 \Big(-405
\left(a^2+r_{+}^2\right)^4
\\\nonumber&\times&\left(a^2+q^2-r_{+}^2\right)-80 \alpha ^2 q^4 \left(15
a^2+26 q^2-15 r_{+}^2\right)+36 \alpha  q^2
\left(a^2+r_{+}^2\right)\\\nonumber&\times&\Big(40 a^4+a^2 \left(51
q^2-40 r_{+}^2\right)+55 q^2 r_{+}^2-80 r_{+}^4\Big)\Big)\Big]
\Big[\Big(9 \left(a^2+r_{+}^2\right)^2\\\nonumber&-&20 \alpha
q^2\Big) \Big(3645 \left(a^2+r_{+}^2\right)^6 \Big(a^4+a^2
\left(q^2+4 r_{+}^2\right)+3 q^2
r_{+}^2-r_{+}^4\Big)\\\nonumber&-&1600 \alpha ^3 q^6 \left(15 a^4+2
a^2 \left(13 q^2+60 r_{+}^2\right)+182 q^2 r_{+}^2-75
r_{+}^4\right)\\\nonumber&-&324 \alpha q^2
\left(a^2+r_{+}^2\right)^3 \Big(65 a^6+76 a^4 q^2+5 r_{+}^4 \left(7
a^2+72 q^2\right)+a^2 r_{+}^2\\\nonumber&\times&\left(425 a^2+404
q^2\right)-325 r_{+}^6\Big)+720 \alpha ^2 q^4
\left(a^2+r_{+}^2\right) \Big(55 a^6+a^4\\\nonumber&\times& \left(77
q^2+435 r_{+}^2\right)+a^2 \left(512 q^2 r_{+}^2+165
r_{+}^4\right)+451 q^2 r_{+}^4\\\label{16}&-&215
r_{+}^6\Big)\Big)\Big]^{-1}.
\end{eqnarray}
Figure \textbf{7} shows that $C$ becomes negative for the considered
domain and hence the system is unstable for small values of the
horizon radius. However, for large values of $r_{+}$, the heat
capacity shows positive trend which yields stable BH solution.
\begin{figure}\center
\epsfig{file=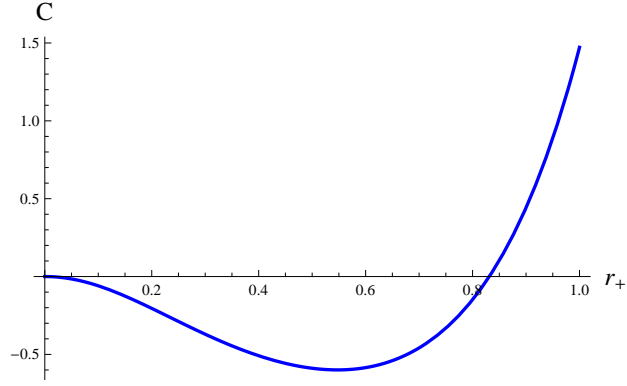,width=0.6\linewidth}\caption{Heat capacity
versus $r_{+}$ for $\alpha=q=0.2$ and $a=0.8$.}
\end{figure}

\section{Null Geodesic and Quasi-normal Modes}

In this section, we discuss null geodesics and photon sphere for the
reduced form of the metric function (\ref{37}). We evaluate Lyapunov
exponent and angular velocity using the photon sphere radius. The
appropriate form of Lagrangian in the equatorial plane
($\theta=0,\frac{\pi}{2}$) is given as \cite{102c}
\begin{equation}\label{28}
2\mathcal{L}=g_{tt}\dot{t}^{2}-\dot{r}^{2}g_{rr}-g_{\phi\phi}\dot{\phi}^{2}.
\end{equation}
The corresponding generalized momentum
($\mathcal{P}_{u}=\frac{\partial\mathcal{L}}{\partial
\dot{x}^{u}}=g_{uv}\dot{x}^{v}$) leads to
\begin{equation}\label{29}
\mathcal{P}_{t}=f(r)\dot{t}\equiv \bar{E},\quad
\mathcal{P}_{\phi}=-R\dot{\phi}\equiv -l,\quad
\mathcal{P}_{r}=-\frac{\dot{r}}{f(r)},
\end{equation}
where $\bar{E}$ and $l$ are the conservation constants which
represent the energy and angular momentum of the photon,
respectively. Using Eq.(\ref{29}), $t$ and $\phi$-motions can be
computed as
\begin{equation}\nonumber
\dot{t}=\frac{\bar{E}}{f(r)},\quad \dot{\phi}=\frac{l}{R}.
\end{equation}
The corresponding Hamiltonian for the null geodesics reads
\begin{equation}\label{30}
2\mathcal{H}=f(r)\dot{t}^{2}-\frac{\dot{r}^{2}}{f(r)}-R\dot{\phi}^{2}
=\bar{E}\dot{t}-l\dot{\phi}-\frac{\dot{r}^{2}}{f(r)}=0,
\end{equation}
leading to
\begin{equation}\label{31}
\dot{r}^{2}+V_{eff}=0,\quad
V_{eff}=\frac{l^{2}}{r^{2}}f(r)-\bar{E}^{2},
\end{equation}
where $V_{eff}$ denotes the effective potential. It is observed that
for $\dot{r}^{2}>0$, the effective potential must be negative which
restricts photon to come out at the region of negative potential.
Thus, the photon will fall into the BH for small value of angular
momentum while for its larger values, the photon will bounce back
before it falls into the BH. Between these cases, there exist
another phase where photon rounds the BH at radial distance with
zero radial velocity \cite{28c4}. These orbits are known as photon
sphere that can be determined through the following conditions
\begin{equation}\label{32}
V_{eff}=0,\quad \frac{\partial V_{eff}}{\partial r}=0,\quad
\frac{\partial^{2} V_{eff}}{\partial r^{2}}<0.
\end{equation}

The first condition leads to the photon sphere radius $(r_{ps})$
while the third condition gives the idea about instability of photon
sphere and links to QNMs of the BH. Inserting Eq.(\ref{31}) into the
second condition, we have
\begin{equation}\label{33}
2f(r_{ps})-r_{ps}f^{'}(r_{ps})=0,
\end{equation}
which, in accordance with Eq.(\ref{37}), gives rise to
\begin{equation}\label{33}
r_{ps}^{4}-3Mr_{ps}^{3}+2q^{2}r_{ps}^{2}-4\alpha q^{2}=0.
\end{equation}
The radius of the photon sphere is given by
\begin{eqnarray}\nonumber
r_{ps}&=&\frac{3 M}{4}+\frac{1}{2}\sqrt{\frac{27 M^3-24 M q^2}{4
\sqrt{\frac{B}{3 A}+\frac{A}{3 \sqrt[3]{2}}+\frac{9 M^2}{4}-\frac{4
q^2}{3}}}-\frac{B}{3 A}-\frac{A}{3 \sqrt[3]{2}}+\frac{9
M^2}{2}-\frac{8 q^2}{3}}\\\label{42}&+&\frac{1}{2} \sqrt{\frac{B}{3
A}+\frac{A}{3\sqrt[3]{2}}+\frac{9 M^2}{4}-\frac{4 q^2}{3}},
\end{eqnarray}
where
\begin{eqnarray}\nonumber
A&=&\Big(-972 \alpha M^2 q^2+\sqrt{\left(-972 \alpha M^2 q^2+576
\alpha q^4+16 q^6\right)^2-4 \left(4 q^4-48 \alpha
q^2\right)^3}\\\label{42a}&+&576 \alpha q^4+16
q^6\Big)^{\frac{1}{3}},\quad B=4 \sqrt[3]{2} \left(q^4-12 \alpha
q^2\right).
\end{eqnarray}
It is noted that for $\alpha=0$, Eq.(\ref{42a}) reduces to RN photon
radius \cite{102c}.

In the eikonal limit $(l\gg1)$, the QNMs can be defined as
\cite{28c5}
\begin{equation}\label{34}
w_{Q}=l\Omega-i\Big(n+\frac{1}{2}\Big)|\lambda|.
\end{equation}
Although this correspondence works for a number of cases
\cite{28CE}, it may violate whenever the perturbations are
gravitational type or the test fields are non-minimally coupled to
gravity \cite{28cCE}. Here $n$ represents the number of overtone,
$\lambda$ and $\Omega$ are the Lyapunov exponent and angular
velocity of the photon sphere given as
\begin{equation}\label{35}
\Omega=\frac{\dot{\phi}}{\dot{t}}\Big|_{r_{ps}}=\frac{f_{ps}}{r_{ps}}
=\frac{1}{l_{ps}},\quad
\lambda=\sqrt{\frac{-V_{eff}^{''}}{2\dot{t}^{2}}}\Big|_{r_{ps}}
=\sqrt{\frac{f_{ps}(2f_{ps}-r_{ps}^{2}f_{ps}^{''})}{2r_{ps}^{2}}}.
\end{equation}
Using Eq.(\ref{37}), these become
\begin{eqnarray}\nonumber
\Omega&=&\frac{\sqrt{\frac{q^2+r_{ps}^2-2 M
r_{ps}}{r_{ps}^2}-\frac{4 \alpha q^2}{3
r_{ps}^4}}}{r_{ps}},\\\nonumber \lambda &=&\sqrt{\frac{\left(12
\alpha q^2-2 q^2 r_{ps}^2+r_{ps}^4\right) \left(3 r_{ps}^2
\left(r_{ps} (r_{ps}-2 M)+q^2\right)-4 \alpha
q^2\right)}{3r_{ps}^{10}}}.
\end{eqnarray}
To analyze the physical behavior, we plot the above expressions
versus charge. It is noted that $\lambda$ shows increasing behavior
from $q=0.53$ as the system is not defined for smaller values of
charge (left plot of Figure \textbf{8}). In the right plot, $\Omega$
depicts negative behavior starting from $q=0.51$ and shows
decreasing trend for large choices of $q$. It is interesting to
mention here that Davies point calculated from Figures \textbf{4}
and \textbf{8} has approximately the same value and lies in the
range $(0.51,0.57)$. The deviation between these values is $6\%$.
\begin{figure}\center
\epsfig{file=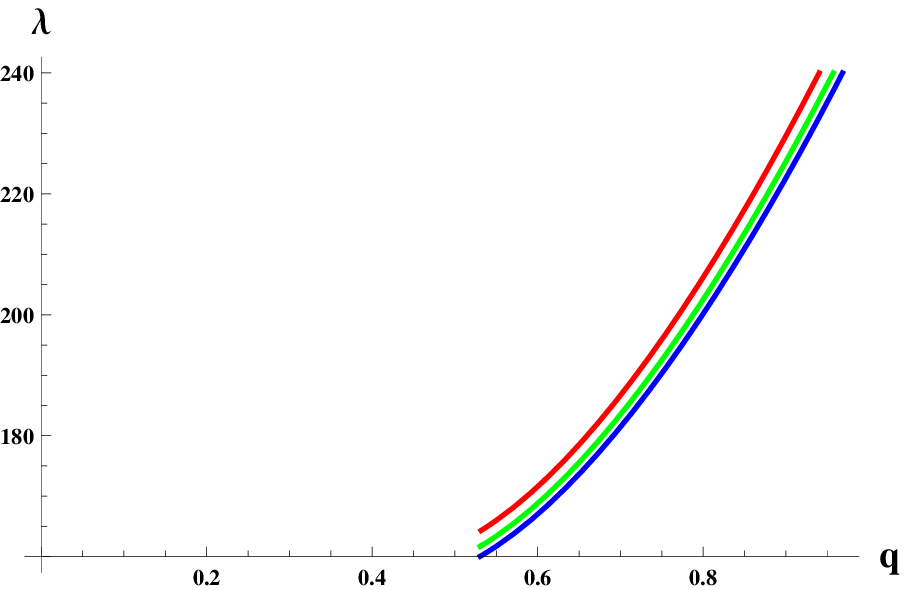,width=0.4\linewidth}
\epsfig{file=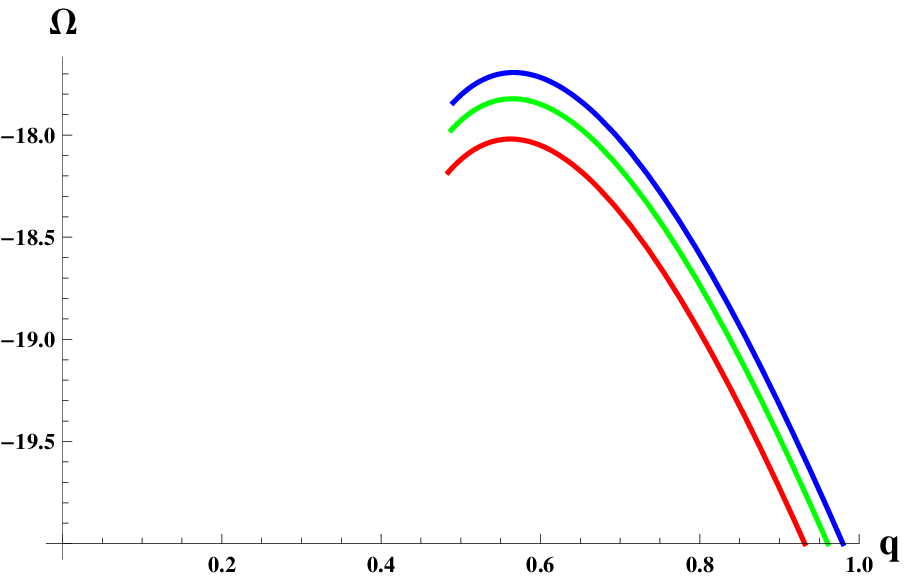,width=0.4\linewidth}
\\\caption{Plots of angular velocity and Lyapunov exponent versus
$q$ for $\alpha=0.02$ (blue), 0.0198 (green) and 0.0195 (red).}
\end{figure}

\section{Thermal Fluctuations}

This section will analyze the effects of thermal fluctuations on
thermodynamic potentials of charged BH with Weyl corrections. The
corrected as well as uncorrected thermodynamic expressions for
entropy, Helmholtz free energy, internal energy, pressure, enthalpy
and specific heat, respectively are computed. Considering
Eq.(\ref{2}) and setting $f(r_{+})=0$, we have
\begin{equation}\label{45}
45r_{+}^{6}-90Mr^{5}+45q^{2}r_{+}^{4}-60\alpha q^{2}
r_{+}^{2}+200\alpha q^{2}Mr_{+}=0.
\end{equation}
The Hawking temperature in terms of $M$ is obtained as
\begin{equation}\label{46}
T=\frac{4 \alpha q^2 \left(5 r_{+} (6 r_{+}-25 M)+78 q^2\right)+45
r_{+}^4 \left(M r_{+}-q^2\right)}{90 \pi r_{+}^7}.
\end{equation}
To investigate the exact expression of entropy against thermal
fluctuations, the partition function is described as \cite{28c..}
\begin{equation}\label{48}
Z(\beta)=\int_{0}^{\infty} dE \rho(E)\exp(-\beta E),
\end{equation}
Using inverse Laplace transform, the density of states is calculated
as
\begin{equation}\label{49}
\rho(E)=\frac{1}{2\pi
i}\int_{\beta_{0}-i\infty}^{\beta_{0}+i\infty}d\beta Z(\beta)
\exp(\beta E) =\frac{1}{2\pi
i}\int_{\beta_{0}-i\infty}^{\beta_{0}+i\infty}d\beta
\exp(\tilde{S}(\beta)).
\end{equation}
where $\tilde{S}(\beta)=\ln Z(\beta)+\beta E$ is known as exact
entropy for the BH which explicitly depends on temperature.
Employing the method of steepest descent, we have
\begin{equation}\label{50}
\tilde{S}(\beta)=S+\frac{1}{2}(\beta-\beta_{0})^{2}
\frac{\partial^{2}\tilde{S}(\beta)}{\partial
\beta^{2}}\Big|_{\beta=\beta_{0}}+\text{(higher-order terms)},
\end{equation}
where $S$ represents equilibrium entropy with $\frac{\partial
\tilde{S}}{\partial\beta}=0$ and $\frac{\partial^{2}
\tilde{S}}{\partial\beta^{2}}>0$. Inserting the above expression in
(\ref{49}), we have
\begin{equation}\label{51}
\rho(E)=\frac{\exp(S)}{2\pi i}\int d\beta
\exp\Big(\frac{1}{2}(\beta-\beta_{0})^{2}
\frac{\partial^{2}\tilde{S}(\beta)}{\partial \beta^{2}}\Big),
\end{equation}
which can further be simplified as \cite{28c...}
\begin{equation}\label{52}
\rho(E)=\frac{\exp(S)}{\sqrt{2\pi}}\Big[\Big(\frac{\partial^{2}\tilde{S}(\beta)
}{\partial
\beta^{2}}\Big)\Big|_{\beta=\beta_{0}}\Big]^{-\frac{1}{2}}.
\end{equation}
Eventually, this leads to
\begin{equation}\label{54}
\tilde{S}=S-\frac{1}{2}\ln(ST^{2})+\frac{\eta}{S}.
\end{equation}
Without loss of generality, we can replace the factor $\frac{1}{2}$
with a more general parameter $\gamma$. In this scenario, the
corrected entropy around thermal equilibrium reads \cite{28cc5}
\begin{equation}\label{55}
\tilde{S}=S-\gamma \ln(ST^{2})+\frac{\eta}{S},
\end{equation}
where $\gamma$ and $\eta$ are correction parameters.
\begin{itemize}
\item For $\gamma,~ \eta \rightarrow 0$, the original BH entropy (entropy without any correction terms)
can be obtained.
\item For $\gamma \rightarrow 1,~ \eta \rightarrow 0$, the usual logarithmic corrections can
be recovered.
\item For $\gamma \rightarrow 0,~\eta \rightarrow 1$, the second
order correction terms can be obtained which is inversely
proportional to original BH entropy.
\item Finally, for $\gamma,~\eta \rightarrow 1$, higher order
corrections can be recovered.
\end{itemize}
\begin{figure}\center
\epsfig{file=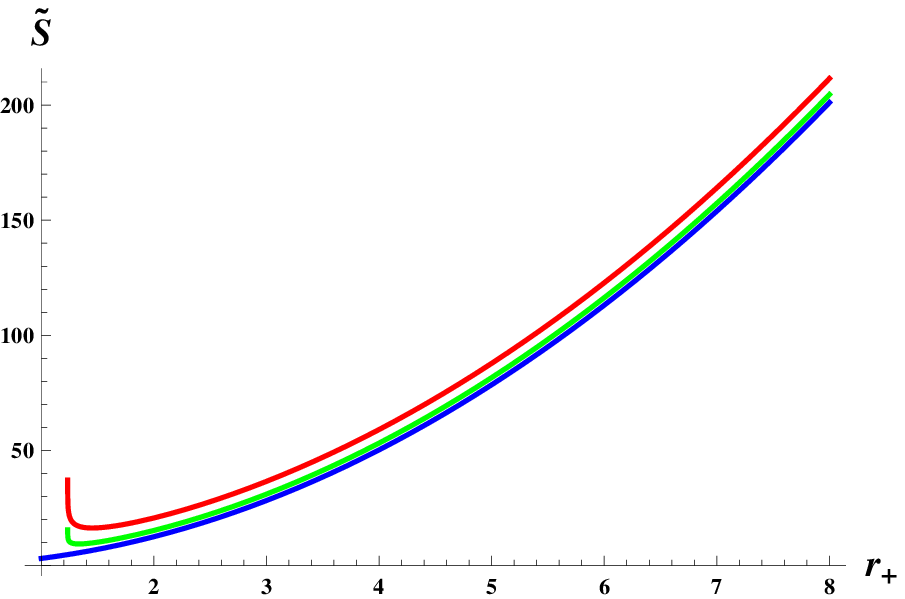,width=0.4\linewidth}
\epsfig{file=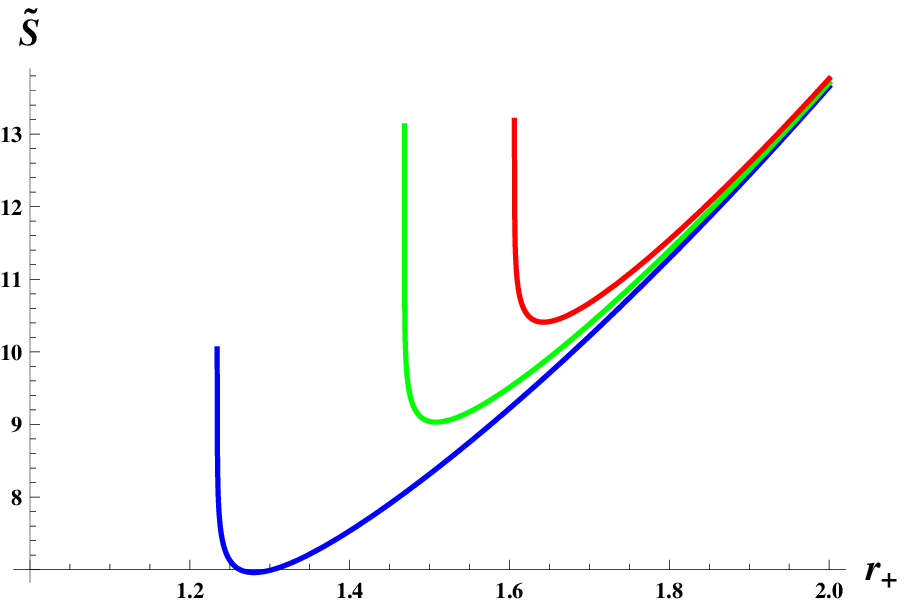,width=0.4\linewidth}\\\caption{Entropy versus
$r_{+}$ for $M=q=1$. We take $\alpha=0.2$ , $\gamma=0$ (blue), 0.5
(green), 1.5 (red) for left plot and $\gamma=0.2$ , $\alpha=0.2$
(blue), 0.6 (green), 1(red) for right plot.}
\end{figure}

Here, we consider the second case $(\gamma \rightarrow 1, ~\eta
\rightarrow 0)$. We see that the second term in the expression
(\ref{55}) is logarithmic in nature which yields the impact of
leading-order corrections to entropy. Inserting Eqs.(\ref{6}) and
(\ref{46}) in (\ref{55}), the perturbed form of entropy becomes
\begin{eqnarray}\nonumber
\tilde{S}&=&\pi r_{+}^2+\gamma \ln \left(8100 \pi
r_{+}^{12}\right)-2 \gamma \ln \Big(4 \alpha q^2 \left(5 r_{+} (6
r_{+}-25 M)+78 q^2\right)\\\label{56}&+&45 r_{+}^4 \left(M
r_{+}-q^2\right)\Big).
\end{eqnarray}
Figure \textbf{9} represents that the entropy of the system remains
positive throughout the considered domain as well as monotonically
increasing for the large BH. The entropy decreases upto a certain
value of the horizon radius which increases gradually for the larger
values of $\gamma$ (left plot) and $\alpha$ (right plot). It is
interesting to mention here that thermal fluctuations are effective
for small BH while the large BHs are unaffected.

In the presence of thermal fluctuations, the modified first law of
BH thermodynamics takes the form \cite{28c}
\begin{equation}\label{56a}
\delta M-\mathcal{T} \delta \tilde{S}- \varphi \delta q-\mathcal{V}
\delta P=0,
\end{equation}
where $\varphi$, $\mathcal{V}$ and $P$ denote the electric
potential, volume and pressure, respectively.  The potential
functions can be obtained through the following relations
\begin{equation}\nonumber
\mathcal{T}=\Big(\frac{\partial M}{\partial
\tilde{S}}\Big)_{q},\quad \varphi=\Big(\frac{\partial M}{\partial
q}\Big)_{ \tilde{S}},\quad \mathcal{V}=\Big(\frac{\partial
M}{\partial P}\Big)_{ \tilde{S}, q},
\end{equation}
with
\begin{eqnarray}\nonumber
\mathcal{T}&=&\Big[4 \Big(\frac{5 \gamma \Big(4 \alpha q^2 (25 M-12
r_{+})+9 r_{+}^3 \Big(4 q^2-5 M r_{+}\Big)\Big)}{4 \alpha q^2 \Big(5
r_{+} (6 r_{+}-25 M)+78 q^2\Big)+45 r_{+}^4 \Big(M
r_{+}-q^2\Big)}+\frac{6 \gamma}{r_{+}}+\pi
r_{+}\Big)\\\nonumber&\times&\Big(20 \alpha q^2 r_{+}-9
r_{+}^5\Big)^2\Big]^{-1}\Big[\alpha^2 \Big(240 q^4 r_{+}^2-416
q^6\Big)+36 \alpha \Big(11 q^4 r_{+}^4-16 q^2
r_{+}^6\Big)\\\nonumber &+&81 r_{+}^8
\Big(r_{+}^2-q^2\Big)\Big],\\\nonumber \varphi&=& \Big[\gamma q
\Big(16 \alpha^2 \Big(26 q^6-15 q^4 r_{+}^2\Big)+\alpha \Big(576 q^2
r_{+}^6-396 q^4 r_{+}^4\Big)+81 r_{+}^8 (q-r_{+})\\\nonumber&\times&
(q+r_{+})\Big)\Big(4 \alpha \Big(5 r_{+} (6 r_{+}-25 M)+156
q^2\Big)-45 r_{+}^4\Big)\Big]\Big[r_{+} \Big(20 \alpha q^2-9
r_{+}^4\Big)^2\\\nonumber&\times& \Big(4 \alpha q^2 \Big(\gamma
\Big(-625 M r_{+}+468 q^2+120 r_{+}^2\Big)+\pi r_{+}^2 \Big(5 r_{+}
(6 r_{+}-25 M)\\\nonumber&+&78 q^2\Big)\Big)+45 r_{+}^4 \Big(\gamma
\Big(M r_{+}-2 q^2\Big)+\pi r_{+}^2 \Big(M
r_{+}-q^2\Big)\Big)\Big)\Big]^{-1}, \\\nonumber
\mathcal{V}&=&\frac{4}{3} r_{+} \Big(-6 \gamma \log \Big(4 \alpha
q^2 \Big(5 r_{+} (6 r_{+}-25 M)+78 q^2\Big)+45r_{+}^4 \Big(M
r_{+}-q^2\Big)\Big)\\\nonumber&+&3 \gamma \log \Big(8100 \pi
r_{+}^{12}\Big)-6 \gamma+\pi r_{+}^2\Big).
\end{eqnarray}
When we substitute the above values in Eq.(\ref{56a}), it is found
that the first law gets satisfied. Thus, it is interesting to
mention here that the logarithmic correction terms increase the
validity of the first law of thermodynamics. Figure \textbf{10}
shows the graphical analysis of corrected temperature which
indicates that the effect of thermal fluctuations is negligible.
Hence, we consider uncorrected temperature along a corrected entropy
to observe the effects of logarithmic corrections.
\begin{figure}\center
\epsfig{file=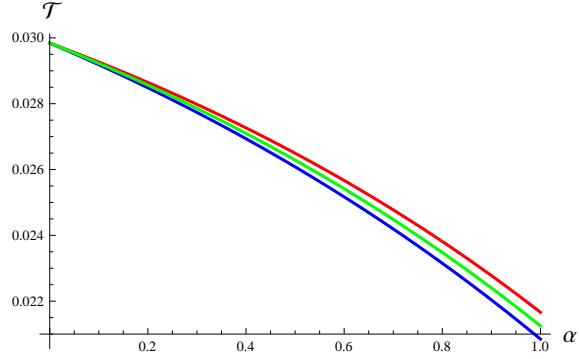,width=0.55\linewidth}\caption{Hawking
temperature versus $\alpha$ for $M=q=1$. We take $r_{+}=2$,
$\gamma=0$ (blue), 0.5 (green), 1.5 (red).}
\end{figure}
\begin{figure}\center
\epsfig{file=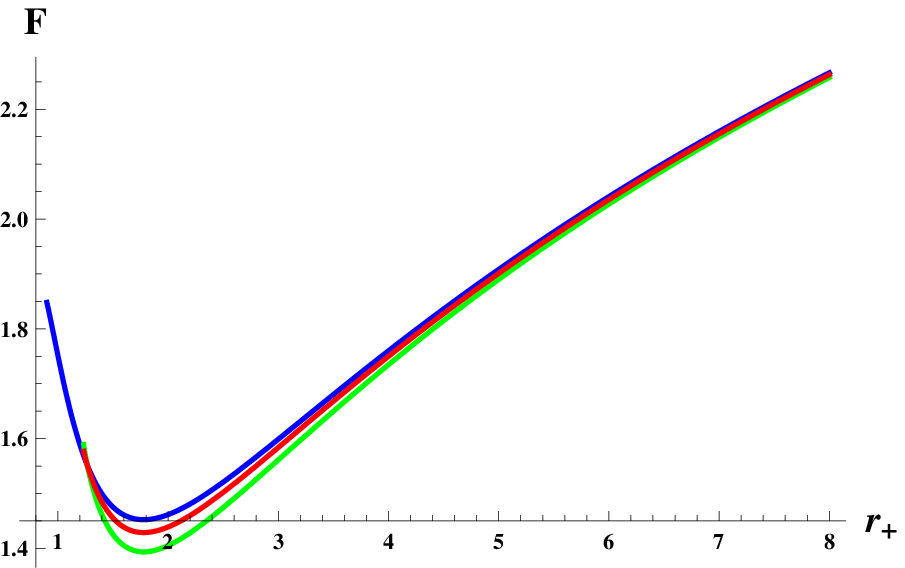,width=0.4\linewidth}
\epsfig{file=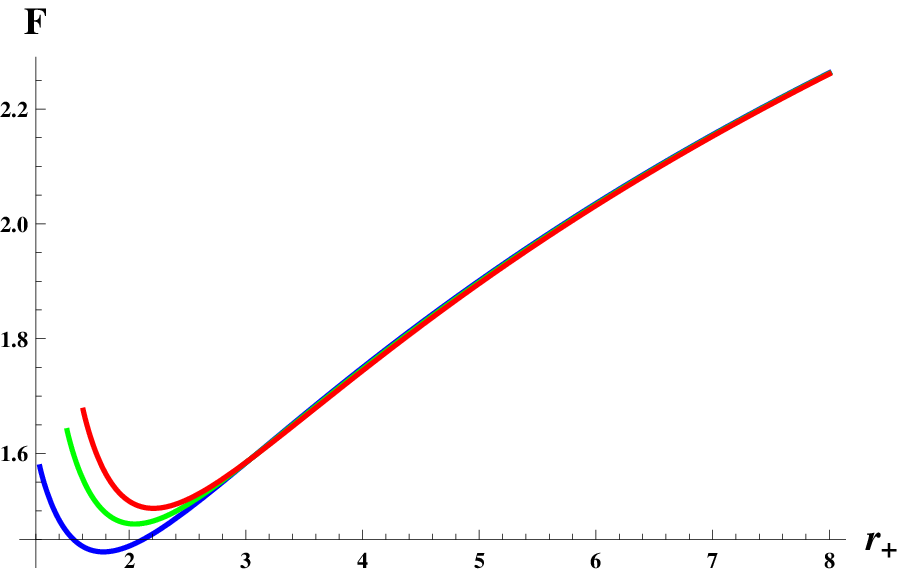,width=0.4\linewidth}\caption{Helmholtz free
energy versus $r_{+}$ for $M=q=1$, $\alpha=0.2$ (left plot) and
$\gamma=0.2$ (right plot).}
\end{figure}

Now, we explore thermodynamical equations of state with the help of
corrected entropy and Hawking temperature. In this respect, the
Helmholtz free energy can be evaluated as
\begin{equation}\label{57}
F=-\int \tilde{S} dT.
\end{equation}
which, through Eqs.(\ref{6}) and (\ref{46}), reduces to
\begin{eqnarray}\nonumber
F&=&-\frac{1}{9450 \pi  r_{+}^7}\Big[2 \alpha q^2 \Big(10 \gamma
\left(7 r_{+} (144 r_{+}-625 M)+2808 q^2\right)+21 \pi  r_{+}^2
\\\nonumber&\times&\left(125 r_{+} (4 r_{+}-15 M)+1092 q^2\right)\Big)+105
\gamma \Big(4 \alpha q^2 \Big(5 r_{+} (6 r_{+}-25 M)\\\nonumber&+&78
q^2\Big)+45 r_{+}^4 \left(M r_{+}-q^2\right)\Big) \Big(\ln
\left(8100 \pi r_{+}^{12}\right)-2 \ln \Big(4 \alpha q^2 \Big(5
r_{+}\\\nonumber&\times& (6 r_{+}-25 M)+78 q^2\Big)+45 r_{+}^4
\left(M r_{+}-q^2\right)\Big)\Big)-1575 r_{+}^4 \Big(-3 \gamma M
r_{+}\\\label{58}&+&4 \gamma q^2+9 \pi q^2 r_{+}^2\Big)-9450 \pi M
r_{+}^7 \ln (r_{+})\Big].
\end{eqnarray}
Figure \textbf{11} gives the behavior of Helmholtz free energy with
respect to $r_{+}$. It is found that the free energy of the small BH
increases corresponding to the larger values of Weyl and correction
parameters while no effect of fluctuations is observed for the large
BH. When the Helmholtz free energy tends to its minimum value, the
system shifts towards its equilibrium state and no further work can
be extracted from it. Thus, the equilibrium condition of maximum
entropy becomes the condition of minimum Helmholtz free energy held
at constant temperature. The internal energy for the considered BH
solution is given by \cite{28c..}
\begin{equation}\label{59}
\mathcal{U}=F + T\tilde{S}.
\end{equation}
Substituting the values of $F$, $T$ and $S$ in the above identity,
it follows that
\begin{eqnarray}\nonumber
\mathcal{U}&=&\frac{1}{9450 \pi r_{+}^7}\Big(-2 \alpha q^2 \Big(10
\gamma \left(7 r_{+} (144 r_{+}-625 M)+2808
q^2\right)\\\nonumber&+&21 \pi r_{+}^2 \left(25 r_{+} (8 r_{+}-25
M)+312 q^2\right)\Big)+1575 r_{+}^4 \\\nonumber&\times&\Big(\gamma
\left(4 q^2-3 M r_{+}\right)+3 \pi r_{+}^2 \left(M r_{+}+2
q^2\right)\Big)\\\label{60}&+&9450 \pi M r_{+}^7 \ln (r_{+})\Big).
\end{eqnarray}
Figure \textbf{12} shows that the internal energy becomes negative
before the critical value of the horizon radius due to thermal
fluctuations whereas for $\gamma=0$, this remains positive
throughout the system. However, after the critical radius, it
observes the same trend as that of Helmholtz free energy.
\begin{figure}\center
\epsfig{file=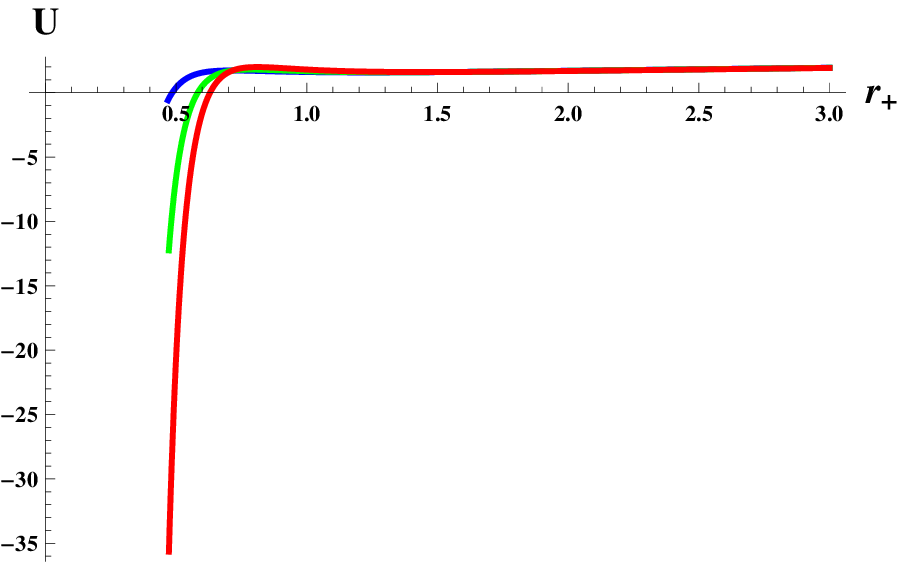,width=0.4\linewidth}
\epsfig{file=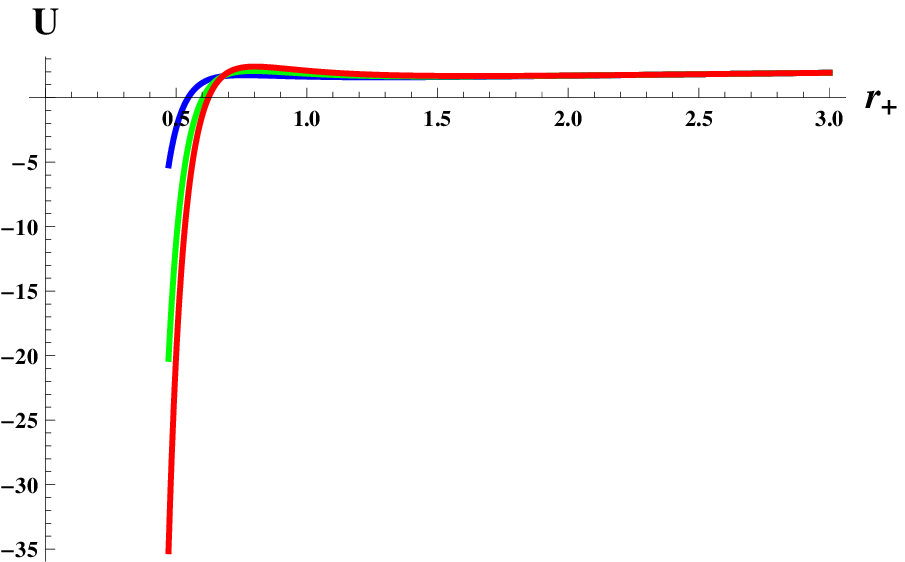,width=0.4\linewidth}\caption{Internal energy
versus $r_{+}$ with same values.}
\end{figure}

The BH volume for considered geometry is defined as \cite{28c6}
\begin{equation}\label{61}
V=\frac{4}{3}\pi r_{+}^{3}.
\end{equation}
Since spacetime is considered as a thermodynamic system, so we
really need to discuss pressure (P). One can calculate BH pressure
using the following relation
\begin{equation}\label{62}
P=-\frac{dF}{dV}=-\frac{dF}{dr_{+}}\frac{dr_{+}}{dV}.
\end{equation}
Using Eqs.(\ref{58}) and (\ref{61}), the pressure takes the form
\begin{eqnarray}\nonumber
P&=&\frac{1}{120 \pi ^2 r_{+}^{10}}\Big[\left(-8 \alpha q^2 \left(25
r_{+} (r_{+}-5 M)+91 q^2\right)-15 r_{+}^4 \left(2 M r_{+}-3
q^2\right)\right) \\\nonumber&\times&\Big(-2 \gamma \ln \left(4
\alpha q^2 \left(5 r_{+} (6 r_{+}-25 M)+78 q^2\right)+45 r_{+}^4
\left(M r_{+}-q^2\right)\right)\\\label{63}&+&\gamma \ln \left(8100
\pi r_{+}^{12}\right)+\pi r_{+}^2\Big)\Big].
\end{eqnarray}
Figure \textbf{13} indicates that the pressure of BH increases
significantly for larger values of the considered parameters and
coincides with the equilibrium pressure for large values of $r_{+}$.
The enthalpy $(H=\mathcal{U}+PV)$ of the system can be obtained as
\begin{eqnarray}\nonumber
H&=&\frac{1}{9450 \pi  r_{+}^7}\Big[-2 \alpha q^2 \Big(10 \gamma
\left(7 r_{+} (144 r_{+}-625 M)+2808 q^2\right)+21 \pi  r_{+}^2
\\\nonumber&\times&\Big(-3125 M r_{+}+2132 q^2+700 r_{+}^2\Big)\Big)-105
\gamma \Big(8 \alpha q^2 \\\nonumber&\times&\left(25 r_{+} (r_{+}-5
M)+91 q^2\right)+15 r_{+}^4 \left(2 M r_{+}-3 q^2\right)\Big)
\Big(\ln \left(8100 \pi r_{+}^{12}\right)\\\nonumber&-&2 \ln \left(4
\alpha q^2 \left(5 r_{+} (6 r_{+}-25 M)+78 q^2\right)+45 r_{+}^4
\left(M r_{+}-q^2\right)\right)\Big)\\\nonumber&+&1575 r_{+}^4
\left(\gamma \left(4 q^2-3 M r_{+}\right)+\pi  r_{+}^2 \left(M
r_{+}+9 q^2\right)\right)\\\label{64}&+&9450 \pi  M r_{+}^7 \ln
(r_{+})\Big].
\end{eqnarray}
The effects of fluctuations on the enthalpy is displayed in Figure
\textbf{14}. It is found that enthalpy of the system is an
increasing function with respect to horizon radius. Moreover, the
corrected as well as equilibrium enthalpy depict the same behavior
for different choices of $\alpha$.
\begin{figure}\center
\epsfig{file=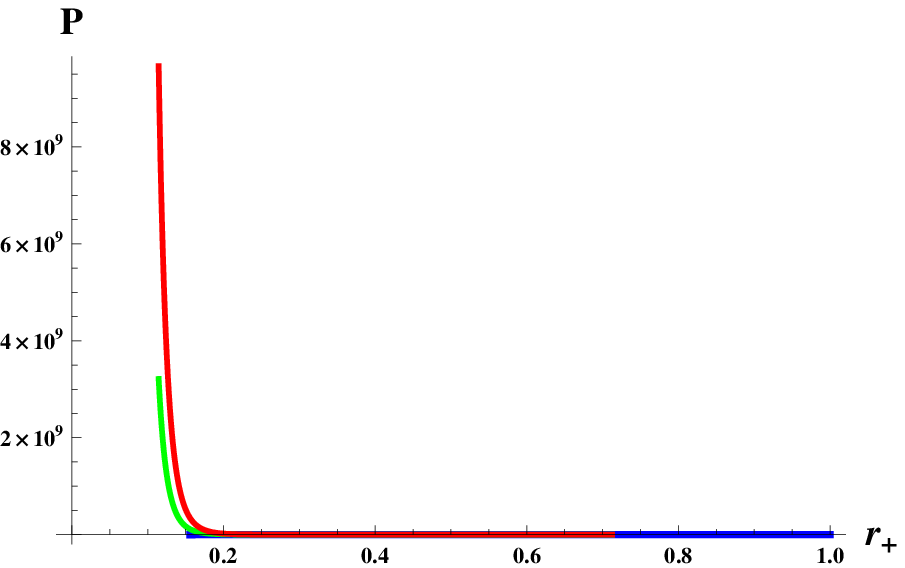,width=0.4\linewidth}
\epsfig{file=
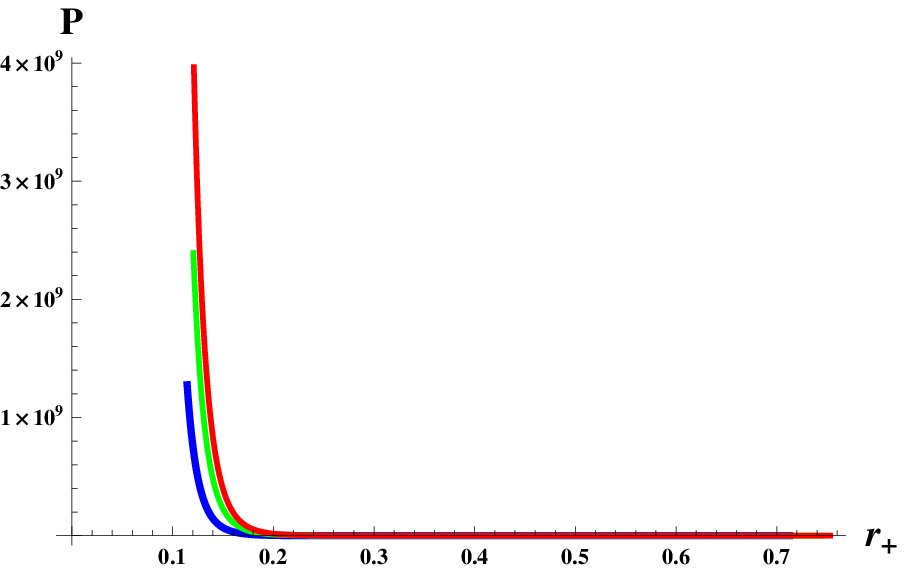,width=0.4\linewidth}\caption{Pressure versus $r_{+}$ with same
parameters.}
\end{figure}
\begin{figure}\center
\epsfig{file=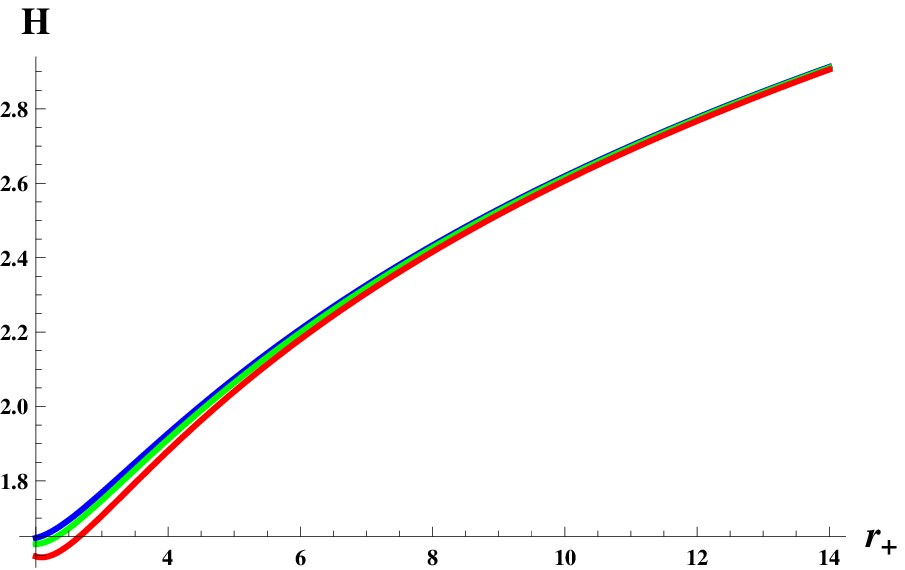,width=0.4\linewidth}
\epsfig{file=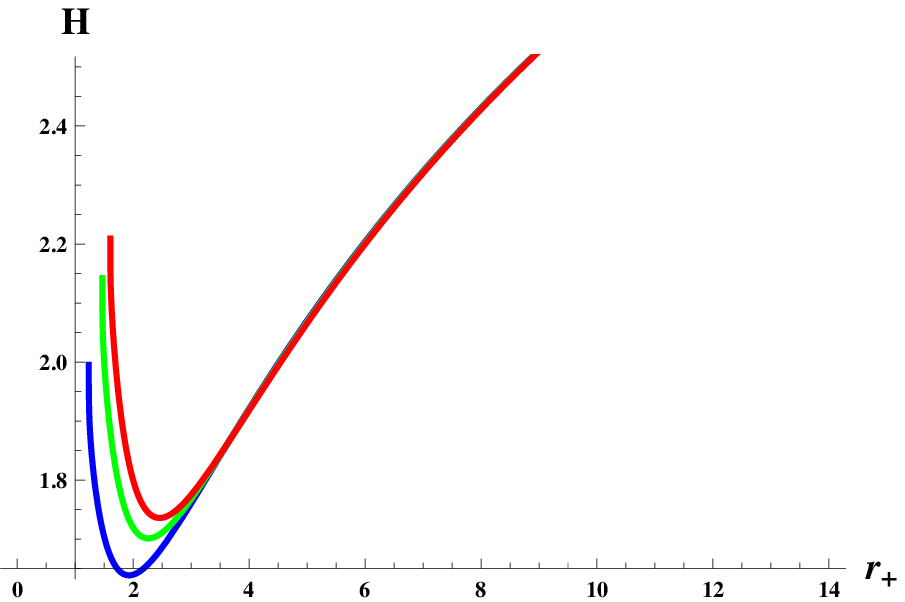,width
=0.4\linewidth}\caption{Enthalpy versus $r_{+}$ with same values.}
\end{figure}
\begin{figure}\center
\epsfig{file=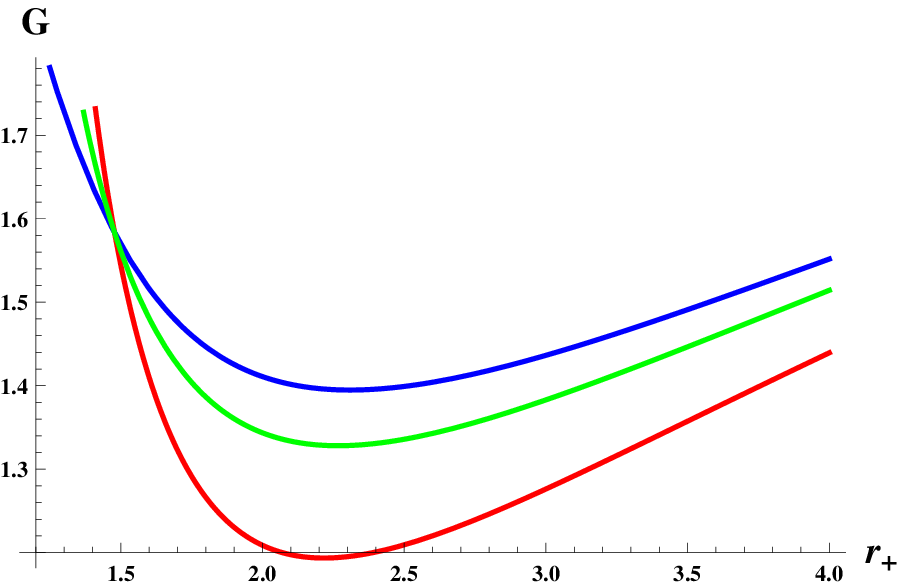,width=0.4\linewidth}
\epsfig{file=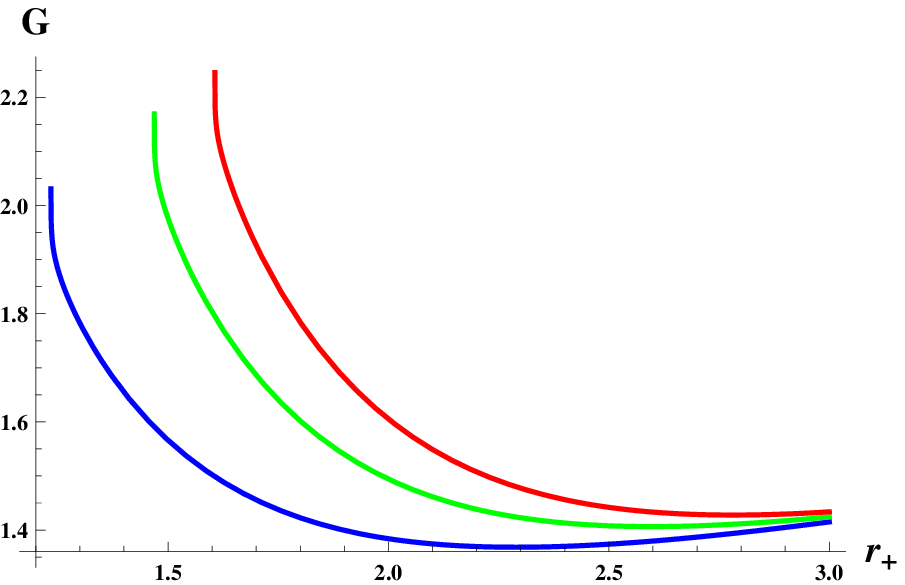,width
=0.4\linewidth}\caption{Gibbs free energy versus $r_{+}$ for
$M=q=1$, $\alpha=0.2$ (left plot) and $\gamma=0.2$ (right plot).}
\end{figure}

The corrected Gibbs free energy $(G=H-T\tilde{S})$ is evaluated as
\cite{28c..}
\begin{eqnarray}\nonumber
G&=&\frac{1}{9450 \pi r_{+}^7}\Big(-2 \alpha q^2 \Big(10 \gamma
\left(7 r_{+} (144 r_{+}-625 M)+2808 q^2\right)\\\nonumber&+&21 \pi
r_{+}^2 \left(125 r_{+} (8 r_{+}-35 M)+2912 q^2\right)\Big)-525
\gamma \Big(4 \alpha q^2\\\nonumber&\times& \Big(-75 M r_{+}+52
q^2+16 r_{+}^2\Big)+3 r_{+}^4 \left(5 M r_{+}-6
q^2\right)\Big)\\\nonumber&\times& \Big(\ln \left(8100 \pi
r_{+}^{12}\right) -2 \ln \Big(4 \alpha q^2\left(5 r_{+} (6 r_{+}-25
M)+78 q^2\right)\\\nonumber&+&45 r_{+}^4 \left(M
r_{+}-q^2\right)\Big)\Big)+1575 r_{+}^4 \Big(\gamma \left(4 q^2-3 M
r_{+}\right)\\\label{66}&-&2 \pi  r_{+}^2 \left(M r_{+}-6
q^2\right)\Big)+9450 \pi M r_{+}^7 \ln (r_{+})\Big).
\end{eqnarray}
Figure \textbf{15} provides that Gibbs free energy decreases against
thermal fluctuations (left plot). The right plot shows the opposite
trend, i.e., Gibbs free energy increases for larger values of Weyl
parameter. In order to examine the stability of charged BH with Weyl
corrections, the specific heat $(C_{S}=\frac{d \mathcal{U}}{dT})$
\cite{28c..} is calculated as
\begin{eqnarray}\nonumber
C_{S}&=&\Big[2 \Big(4 \alpha q^2 \Big(\gamma \left(-625 M r_{+}+468
q^2+120 r_{+}^2\right)+\pi  r_{+}^2 \Big(5 r_{+} (6 r_{+}-25
M)\\\nonumber&+&78 q^2\Big)\Big)\\\nonumber&+&45 r_{+}^4
\left(\gamma \left(M r_{+}-2 q^2\right)+\pi r_{+}^2 \left(M
r_{+}-q^2\right)\right)\Big)\Big]\Big[3 \Big(15 r_{+}^4 \left(3
q^2-2 M r_{+}\right)\\\label{68}&-&8 \alpha q^2 \left(25 r_{+}
(r_{+}-5 M)+91 q^2\right)\Big)\Big]^{-1}.
\end{eqnarray}
Figure \textbf{16} represents that specific heat diverges at
$r_{+}=0.8$ indicating the phase transition of charged BH. We note
that specific heat is negative before the phase transition which
shows that small BHs are unstable under the fluctuations while
becomes stable after phase transition as heat capacity attains
positive values.
\begin{figure}\center
\epsfig{file=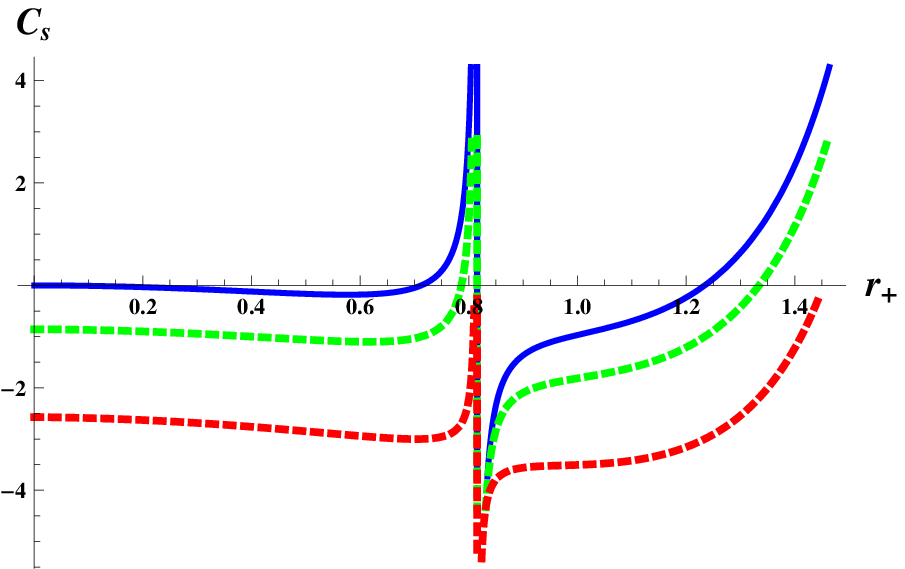,width=0.4\linewidth}
\epsfig{file=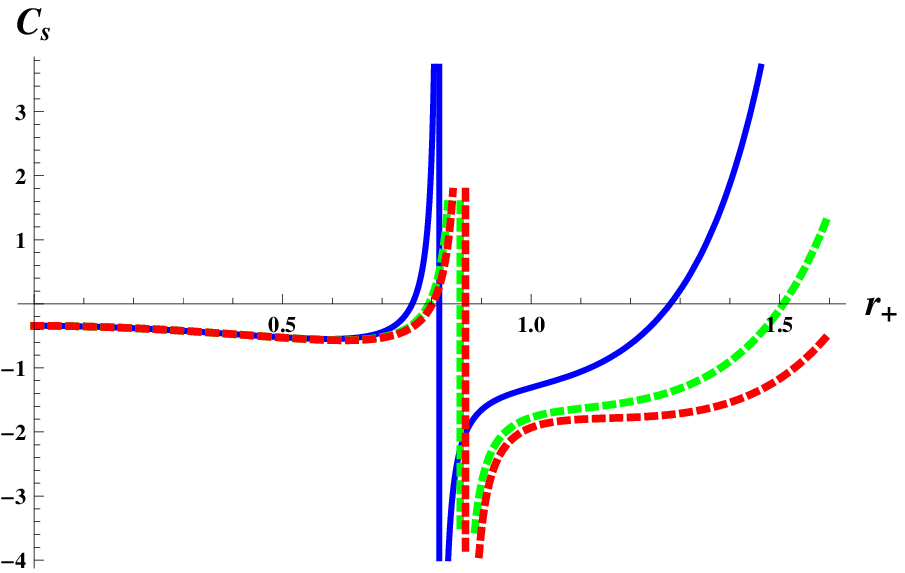,width=0.4\linewidth}\caption{Specific heat versus
$r_{+}$ for $M=q=1$, $\alpha=0.2$ (left plot) and $\gamma=0.2$
(right plot).}
\end{figure}

We can check stability of the system with the help of Hessian matrix
which contains second derivatives of Helmholtz free energy with
respect to temperature $T$ and chemical potential
$\upsilon=\Big(\frac{\partial M}{\partial q}\Big)_{r_{+}}$. The
Hessian matrix is given by \cite{28'}
\begin{equation*}\label{1}
H=\left(
\begin{array}{cc}
H_{11} & H_{12} \\
H_{21} & H_{22} \\
\end{array}
\right),
\end{equation*}
where
\begin{equation}\nonumber
H_{11}=\frac{\partial^{2} F}{\partial T^{2}},\quad
H_{12}=\frac{\partial^{2} F}{\partial T \partial \upsilon},\quad
H_{21}=\frac{\partial^{2} F}{\partial \upsilon \partial T },\quad
H_{22}=\frac{\partial^{2} F}{\partial \upsilon^{2}}.
\end{equation}
The determinant of matrix implies that one of the eigenvalues is
zero as $H_{11}H_{22}=H_{12}H_{21}$. Thus, we use trace of the matrix
to determine the stability given by
\begin{equation}\nonumber
\tau\equiv Tr(H)=H_{11}+H_{22}.
\end{equation}
A necessary criterion for the stable spacetime is the positivity of
trace of the Hessian matrix, i.e., $Tr(H)\geq0$ \cite{28c7e}. From
the graphical analysis of trace with respect to horizon (Figure
\textbf{17}), it is clear that small black holes fulfill the
stability criterion while the large ones depict unstable behavior.
Hence, we find that the Weyl and logarithmic corrections
significantly affect the critical point as well as stability of
small BHs.
\begin{figure}\center
\epsfig{file=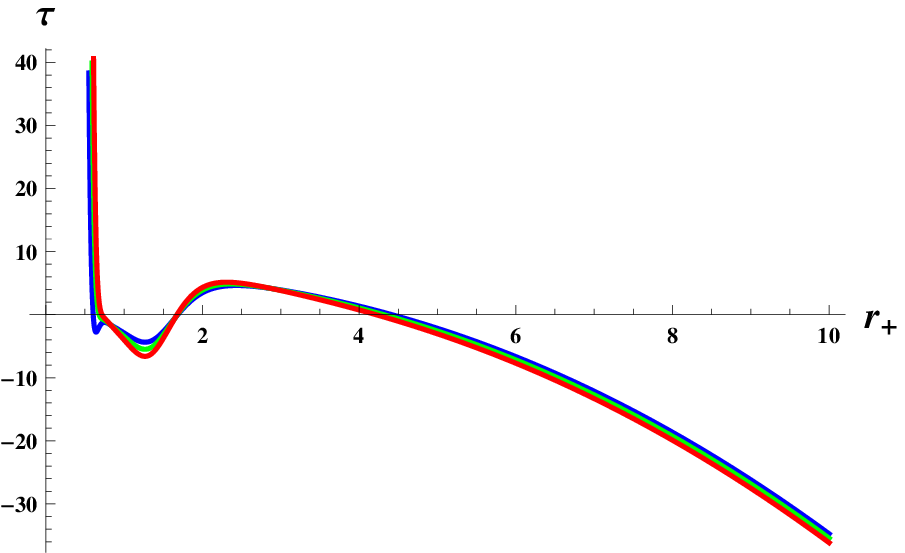,width=0.4\linewidth}
\epsfig{file=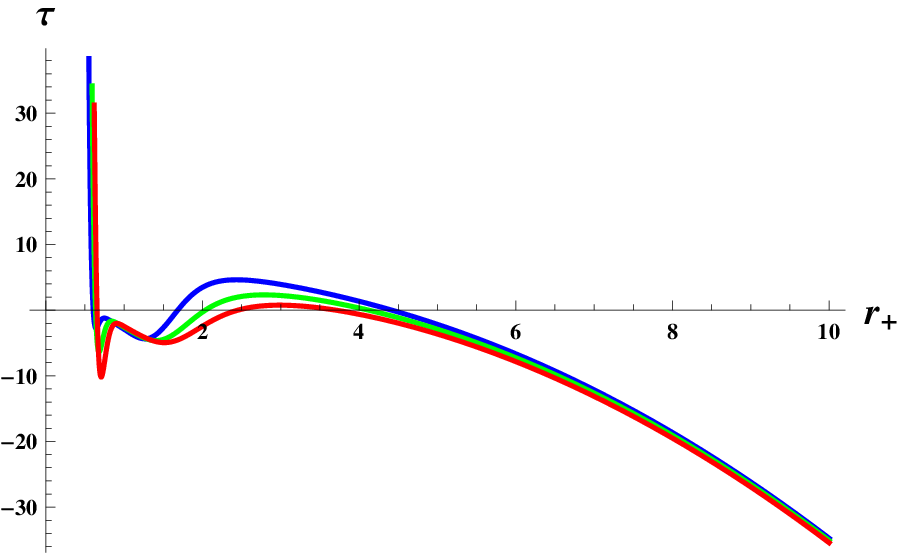,width=0.4\linewidth}\caption{Hessian trace versus
$r_{+}$ for $M=q=1$. We take $\alpha=0.2$ , $\gamma=0.5$ (blue),
-0.5 (green), 0 (red) for left plot and $\gamma=0.2$, $\alpha=0.5$
(blue), -0.5 (green), 0 (red) for right plot.}
\end{figure}

\section{Concluding Remarks}

In this paper, we have studied thermodynamic quantities, QNMs as
well as logarithmic corrections for non-rotating charged BH with
Weyl corrections. Firstly, the Hawking temperature is computed
through surface gravity as well as quantum tunneling and obtained
the same result. We have then used null geodesics as well as photon
sphere radius to derive relation between Davies point and QNMs. We
have also discussed the effect of logarithmic corrections and
compared the results of corrected as well as uncorrected
thermodynamic potentials through graphical analysis. We have found
that heat capacity diverges at $r_{+}=0.27, 0.72$ and attained
positive values for small charged BH (non-rotating) but BH
(rotating) is stable for larger values of horizon radius. It is
noted that the Weyl coupling parameter decreases the temperature of
BH for both rotating as well as non-rotating scenarios. We have
observed that Hawking temperature of non-rotating charged BH is
slightly larger than the rotating (Figures \textbf{2} and
\textbf{6}).

Secondly, we have investigated the relation between QNMs and Davies
point which provides real as well as imaginary parts of QNMs as
angular velocity and Lyapunov exponent, respectively. It is shown
that the Lyapunov exponent increases for larger values of charge
while angular velocity shows opposite behavior for the considered
domain. We would like to mention here that Davies points calculated
from heat capacity and QNMs have approximately the same values,
i.e., $0.57$ and $0.51$, respectively with $6\%$ deviation. This is
less than the deviation measured for the charged BH without
correction parameter \cite{102c}.

Finally, we have considered the first-order logarithmic corrections
to entropy that modify all given thermodynamic potentials except
temperature which is similar to isothermal process. It is seen that
entropy is positive valued function and shows decreasing
(increasing) behavior for the smaller (larger) values of $r_{+}$.
For small horizon radius, the Helmholtz free energy decreases
corresponding to larger fluctuation parameter and coincides with the
equilibrium state for large values of $r_{+}$ which shows that
thermal fluctuations only affect the small BH geometries. The
internal energy becomes negative for the smaller values of horizon
radius which implies that the temperature of the BH falls due to
thermal fluctuations.

The behavior of Gibbs free energy is positive and shows a decreasing
trend for larger values of $\gamma$ which indicates that reactions
occur inside the BHs are non-spontaneous, i.e., BH requires external
energy to sustain its equilibrium position. It is found that
specific heat is negative for small BH indicating unstable phase
while the system is stable for large values of horizon radius in the
presence of thermal fluctuations. Hence, thermal fluctuations induce
more instability for small and medium BHs while the large BHs remain
unaffected. It is worth mentioning here that for $\alpha=0$, all the
derived results reduce to RN BH \cite{28ci,29b}.

\end{document}